# Real-Time Data Processing in the Muon System of the D0 Detector

N. Parashar [5], O. Bardon [5], R. Goodwin [4], S. Hansen [4], B. Hoeneisen [1,2], V. Podstavkov [3], S. Uvarov [3], D. Wood [5]

1. Universidad de los Andes, Bogota, Colombia
2. Universidad San Francisco de Quito, Quito, Ecuador
3. Petersburg Nuclear Physics Institute, St. Petersberg, Russia
4. Fermi National Accelerator Laboratory, Batavia, Illinois 60510, USA
5. Northeastern University, Boston, Massachusetts 02115, USA

*Abstract*

This paper presents a real-time application of the 16-bit fixed point Digital Signal Processors (DSPs), in the Muon System of the D0 detector located at the Fermilab Tevatron, presently the world's highest-energy hadron collider. As part of the Upgrade for a run beginning in the year 2000, the system is required to process data at an input event rate of 10 KHz without incurring significant deadtime in readout.

The ADSP21csp01 processor has high I/O bandwidth, single cycle instruction execution and fast task switching support to provide efficient multisignal processing. The processor's internal memory consists of 4K words of Program Memory and 4K words of Data Memory. In addition there is an external memory of 32K words for general event buffering and 16K words of Dual Port Memory for input data queuing. This DSP fulfills the requirement of the Muon subdetector systems for data readout.  All  error  handling, buffering, formatting and transferring of the data to the various trigger levels of the data acquisition system is done in software. The algorithms developed for the system complete these tasks in about 20 µs per event.

## I. INTRODUCTION

The D0 detector at the Fermilab Tevatron has three subsystem detectors designed for the Muon Upgrade Project [1,2]. All the three subsystems, namely PDTs (Proportional Drift Tubes), MDTs (Mini-Drift Tubes) and Scintillators interact with two levels of triggers. All subsystems participating in the trigger send trigger data to the Trigger Framework (TFW) which is the source of all trigger decisions. These decisions are made on two levels based on inputs from the Level 1 (L1) and Level 2 (L2) trigger systems.  L1 generates trigger information synchronously with the beam crossings while L2 operates asynchronously with an indeterminate decision time within some limits.

A major goal achieved is that all  three detector subsystems follow a common strategy for readout, which is based on a 16-bit fixed point digital signal processor (DSP), to perform readout, buffering and formatting of the data and eventually to transfer them to the trigger systems [3]. The Front-End Units are controlled by Control Units, each of which is equipped with a DSP. All front-end channels (PDTs, MDTs or Scintillators) are equipped with test pulse inputs, with the pulses themselves being generated by the Front-End Units. The enabling, delays and amplitudes of the pulsers are controlled from the Control Units. Table I describes the salient features of the three subdetectors of the Muon System. The ADSP-21csp01 processor operates at 50 MHz (20 ns/instruction), fast enough to execute the multiple algorithms and perform the multiple tasks of the application in real-time. It addresses a large memory space of up to 16M words of instructions and data to meet the requirements of all the subsystems. In addition, it has multiple I/O ports and a Direct Memory Access (DMA) channel to stream the data out of the DSPs internal memory without interrupting the processor. Another important feature of processing multiple signals efficiently in real-time is the 'interrupt latency'. The ADSP-21csp01 responds to external and internal interrupts in as little as 60 ns (3 cycles).

The primary purpose of the DSP is to buffer the data from the Front-End, while a Level 2 decision is pending. The secondary purpose is to re-format this data, if accepted by L2 and send it to the Level  3 trigger system, which   is a bank of NT workstations running a software filter. A check on the integrity of the data is performed at every stage of processing. Figure 1 shows the buffering scheme for the Muon System.

## II. DSP DATA PROCESSING

Continuously running digitizers operate synchronously with the Tevatron RF. The digitizer outputs are connected to pipeline delays, running synchronously with the digitizers to store the data, long enough  to make an L1 decision in about 4.8 µs. When the L1Accept signal arrives at the front-end, the corresponding event information appears at the pipeline output and is transferred to a L1 FIFO. From this point onward, the DSP takes over. For Run II, the L1 trigger rate is 10 KHz and the L2 trigger rate is 1KHz.  All L1 events are sent to L2, which accepts only 10 %. In other words, an average interval of 100 µs is available for transferring an event from the output of the L1 delay to the L2 trigger system and the L2 decision time. The Research Queuing Package (RESQ) from IBM was used to study the expected deadtime due to the L2 trigger. The RESQ model predicts a deadtime of 0.2% for the system. Based on this queuing simulation of the D0 data acquisition system, the L2 latency from the Muon Front-Ends cannot exceed 30 µs, without introducing additional deadtime.

Table 1: The three subdetectors of the Muon System.

| PDT | MDT | Scintillators |
|---|---|---|
| 9, 500 channels | 50,000 channels | 6,000 channels |
| 24-channel Front-End Boards (FEBs) | 192-channel 9U VME Mini-drift tube Readout Controllers | 48-channel 9U VMEScintillator Front-End Cards (SFEs) |
| 3-4 FEBs attached to one Control Board (CB) | Mini-drift tube Readout Controllers (MDRCs) | Scintillator Readout Controllers (SCRCs) |
| 1 DSP for each CB | 1 DSP for each MDRC | 1 DSP for each SCRC |
| Total DSP no is 94 | Total DSP no is 24 | Total DSP no is 18 |

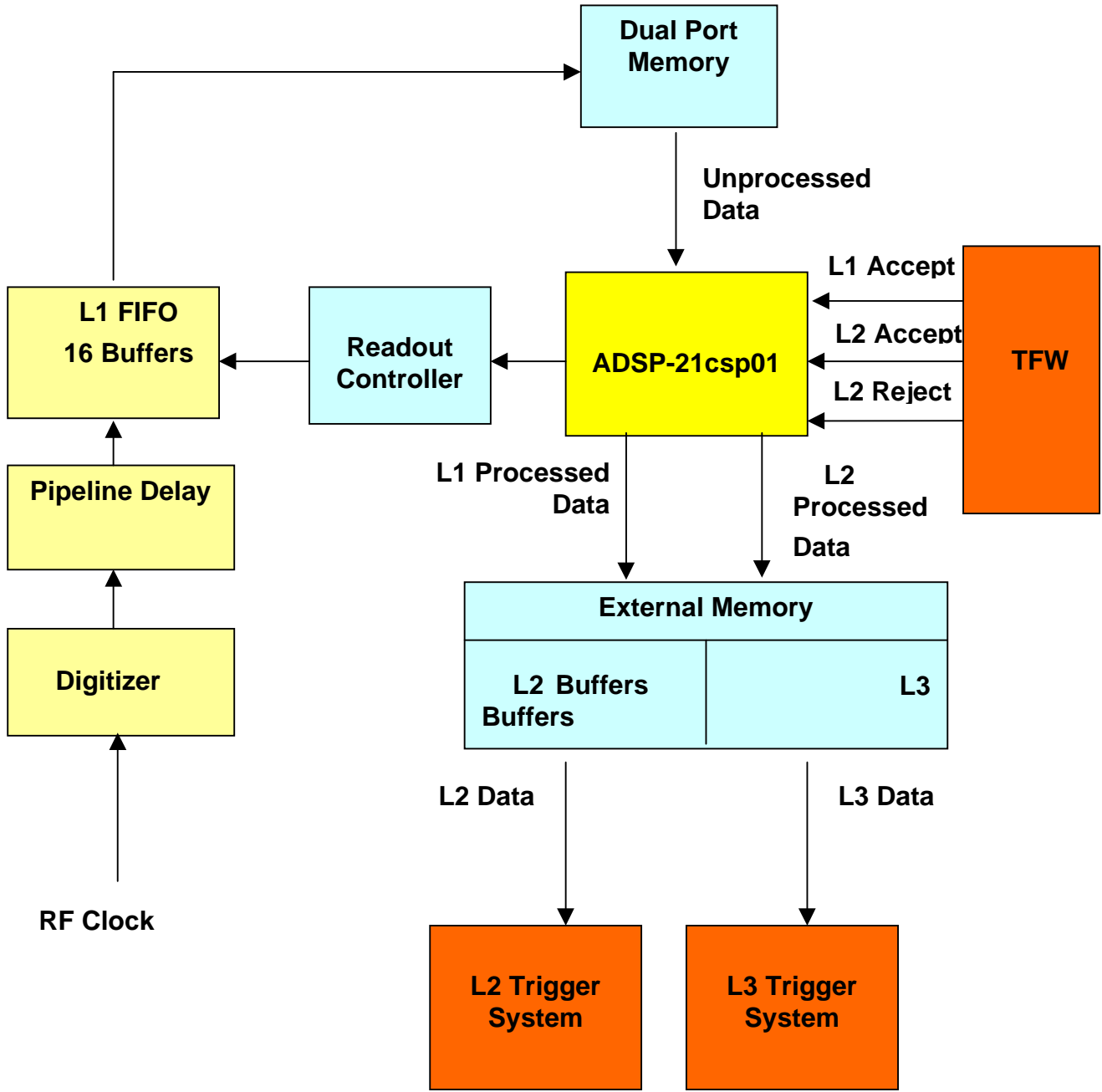

Figure 1: D0 Muon Buffering Scheme

Table 2: L2 processing times for the three subdetectors of the Muon System.

| Detector System | Occupancy (%) | No. of Data words | T1 (µs) | T2 (µs) | T3 (µs) | T4 (µs) | Total L2 Latency (µs) |
|---|---|---|---|---|---|---|---|
| PDT | 3.0 (E) | 6 | 1.5 | 5.0 | 10.0 | 1.7 | 18.2 |
| | 5.0 (C) | 10 | 1.5 | 5.1 | 11.8 | 2.1 | 20.5 |
| MDT | 0.5 (E) | 14 | 0.1 | 3.0 | 7.2 | 5.0 | 25.3 |
| | 1.0 (C) | 27 | 0.1 | 13.0 | 10.6 | 8.0 | 31.7 |
| Scintillator | 1.0 (E) | 5 | 0.1 | 10.6 | 5.0 | 2.1 | 17.8 |
| | 2.0 (C) | 10 | 0.1 | 11.9 | 7.0 | 3.4 | 22.4 |

The processor's internal (on-chip) program memory of 4K is used to load the instructions for the execution of the code, and to prepare the 16 L2 and 8 L3 header block information buffers. The internal data memory of 4K is used to store the variables and lookup tables needed during code execution, and to copy one event data at a time into a buffer of size 1K. A crucial requirement imposed by the system is the minimum interval of 2.6 µs between two L1 triggers. The DSP must process L1 Accepts in less than 2.6 µs to avoid missing any further incoming L1 Accepts. On receipt of an L1 Accept Interrupt, the processor starts a Readout Controller which copies the data from the L1 FIFOs to a Dual Port Memory (DPM), which is 16 buffers deep, each of size 1K (1024 words). The processor here checks for event synchronization by comparing the beam crossing numbers coming from the TFW and the local counter on the Control Unit, and sets the busy and error flags if needed. This preliminary task is completed in less than 1 µs after which the DSP is ready to take another L1 event, thereby abiding by the 2.6 µs constraint.

When the event data has been transferred to the DPM, the processor gets a Read_Done Interrupt. The data is copied on-chip, sorted and converted into physical units of arrival time of particle hits in ns and drift distance in µm, according to the requirements of the L2 trigger system. The formatted L2 event is copied to external memory (32K words), in a buffer of size 1K and is transferred to the L2 trigger system from there via the DMA. An internal interrupt indicates that the data transfer is complete.

The DSP receives a decision from the L2 system in the form of an L2 interrupt. The event synchronization is checked and the event Accept/Reject flag is read. An L2 reject frees a buffer in the DPM, since this event does not need to be sent to L3. An L2 Accept on the other hand, causes the DSP to copy the data from the DPM into a buffer in the external memory. There are 8 such buffers, each of size 1K, to hold the event raw data. Following this, the corresponding event buffer in the DPM is freed to allow one more event buffering. The copied event data is then processed in the background, according to the requirements of L3 system and queued into a buffer in the external memory. There are 8 of these buffers, each of size 1K, to store the formatted L3 data. The DSP then waits for the permission to transfer L3 data to the Muon Readout Card (MRC) for final event building in the muon readout crate. The L2 accept rate of 1 KHz is input to L3, i.e. the background formatting, processing and transmission to L3 should be done within 1 ms.

The DSP code, written in Assembly language, handles the various interrupts in order of priority, and deals with the error-checking conditions, as a requirement of the system. These are listed below in order of priority, with an outline of the tasks they perform.

1. L1 Interrupt (Highest priority)
   - Checks the availability of buffers.
   - Starts the Readout Controller.
   - Compares Local and TFW L1 Crossing Numbers.
2. L2 Accept/Reject Interrupt
   - Compares Local and TFW L2 Crossing Numbers.
   - Checks L2 Accept/Reject flag.
   - If Reject, discards the buffer in the DPM.
   - If Accept, copies the data from the DPM to be prepared for L3 in the background, frees up the DPM buffer.
   - Set L2 Busy and L2 Error, if necessary.
3. Internal DMA Interrupt
   - Clear internal DMA busy flag.

4. Read_Done Interrupt (Lowest priority)
- Reads data from the DPM.
- Processes and formats the data for L2.
- Transmits the formatted data to L2.

5. Background process
- Processes and formats the data for L3.
- Transmits the formatted data to L3.

The readout times depend upon the size of the event and the complexity of the algorithms. The L2 processing times for the expected (E) and conservative (C) occupancies for the PDTs, MDTs and Scintillators are listed in Table 2. The L2 latency =T1+T2+T3+T4, where,

T1: Time taken to transfer the L1 Accepted data into L1 FIFOs, since the trigger generated.

T2: Time taken to transfer the data from the L1 FIFOs to the DPM.

T3: The L2 processing times in the Readout DSPs.

T4: The L2 data transmission time from the DSP to the L2 trigger system.

## III. CONCLUSIONS

The crucial result of this work is that there is no significant deadtime introduced in the readout at the expected occupancy rates in the new environment for a run in the year 2000, operating at an input rate of 10 KHz. The ADSP-21csp01 processor performs all necessary time-critical formatting, buffering, error-handling and data transmission as required by the system in about 20 $\mu$s at the expected occupancy rates, thereby respecting the 30 $\mu$s output limit from this DSP. The software is now being tested on a prototype readout chain. The code responds to the various interrupts and hardware flags. The final Electronics Readout Configuration is required to test the code completely.

## IV. ACKNOWLEDGMENTS

We thank the staffs at Fermilab and collaborating institutions for their contributions to this work, and acknowledge support from the Department of Energy and National Science Foundation (U.S.A.), Ministry for Science and Technology and Ministry for Atomic Energy (Russia) and Colciencias (Colombia).